\documentclass[aps,prl,twocolumn,superscriptaddress,]{revtex4}
\usepackage{setspace} 
\usepackage{amsmath}
\usepackage{graphicx}
\usepackage{mathrsfs}
\usepackage{bm}        
\usepackage{longtable}
\usepackage{tabularx}
 \usepackage{bm}        
\usepackage{amssymb}   
\usepackage{amsmath}
\usepackage{color}
\usepackage[toc,page]{appendix}
\usepackage{mathtools}

\def\bea{\begin{eqnarray}}
\def\eea{\end{eqnarray}}

\def\p{\partial}

\def\la{\langle}
\def\ra{\rangle}

\begin{document}


\title{Modelling of \textit{Dictyostelium discoideum} movement in linear gradient of chemoattractant }

\author{Zahra Eidi}
\affiliation{Department of Physics, Institute for Advanced Studies in
Basic Sciences (IASBS), Zanjan 45137-66731, Iran}
\author{ Farshid Mohammad-Rafiee} 
\affiliation{Department of Physics, Institute for Advanced Studies in
Basic Sciences (IASBS), Zanjan 45137-66731, Iran}
\author{Mohammad Khorrami}
\affiliation{	Department of Physics, Alzahra University, Tehran, Iran} 	
\author{Azam Gholami}
\affiliation{	Max Planck Institute for Dynamics and Self-Organization,
	 G\"{o}ttingen, Germany
}

\date{\today}

\begin{abstract}

Chemotaxis is a ubiquitous biological phenomenon in which cells detect a spatial gradient of chemoattractant, and then move towards the source. Here we present a position-dependent advection-diffusion model that quantitatively describes the statistical features of the chemotactic motion of the social amoeba {\it Dictyostelium discoideum} in a linear gradient of cAMP (cyclic adenosine monophosphate).  We fit the model to experimental trajectories that are recorded in a microfluidic setup with stationary cAMP gradients and extract the diffusion and drift coefficients in the gradient direction. Our analysis shows that for the majority of gradients, both coefficients decrease in time and become negative as the cells crawl up the gradient. The extracted model parameters also show that besides the expected drift in the direction of chemoattractant gradient, we observe a nonlinear dependency of the corresponding variance in time, which can be explained by the model. Furthermore, the results of the model show that the non-linear term in the mean squared displacement of the cell trajectories can dominate the linear term on large time scales. 

\end{abstract}

\maketitle

\section{I. Introduction}

\textit{Dictyostelium discoideum} ({\it D.d.}) is a well-established model organism for cellular motility. Chemotactic competent {\it D.d.}  cells are highly motile and exhibit fast amoeboid movements with a velocity of $10-20~\mu m/min$ on glass substrates~\cite{Takagi-2008,Loomis:2012,Sander-2010}. The chemotactic cell motion is highly organized over a length scale significantly larger than the size of a single cell ($\sim$10 $\mu$m). When nutrients are depleted, {\it D.d.} cells secret a chemical called cAMP (cyclic adenosine monophosphate) that  has  an  attractive  effect  on  the  cells  themselves. Cells sense gradients of cAMP and direct their chemotactic movements towards regions of higher concentration of cAMP \cite{Devreotes-2004}.  When chemotactic attraction prevails over diffusion, the chemotaxis can  trigger  a  self-accelerating  process  until aggregation takes place.  As a result, $10^5$ -- $10^6$ cells stream towards the aggregation centers and eventually transform into millimeter long slugs and ultimately form fruiting bodies bearing spores for long-term survival and long-range dispersal \cite{book-Kessin}.

Different mathematical models incorporate chemotaxis in different ways; however, a common mechanism is to assume that chemotaxis biases the otherwise random motion of crawling cells along the concentration gradients of chemoattractants~\cite{Patlak-1953}. The random cell movement is commonly described as a diffusion and the directional movement along the chemical gradient is incorporated as a combination of diffusion and advection. In the simplest model, the diffusion coefficient and the drift velocity of the cells are assumed to be constant. However, in general, these coefficients depend on both the absolute concentration and the gradient of the chemical~\cite{book-Friedman-Kao,Segel-1970,Amselem-2012,Amselem-2012-PRL}. The advection-diffusion equation has been previously used to describe the aggregation phase of {\it D.d.} cells where the chemotactic force pulls the amoebas towards the aggregation centers. For example, a model of slime mold aggregation has been introduced by Patlak~\cite{Patlak-1953} and Keller~\cite{Segel-1970} in the form of two coupled differential equations. The first equation is an advection-diffusion equation describing the evolution  of  the  concentration of  amobae  and  the  second equation is a diffusion equation with terms of source and
degradation describing the evolution of the concentration of the signaling molecule. The original form of the Keller-Segel model, would allow the diffusion coefficient and the drift velocity to depend on the cAMP concentration and on the concentration of the amoeba. The case that these coefficients depend on the chemical cencentration but not on the cell density has been considered by Othmer and Stevens~\cite{Othmer-1997}. This leads to ordinary mean field Fokker-Planck equations for cell density with space and time dependent coefficients~\cite{Chavanis-2008}. On the other hand, if we assume that the diffusion coefficient and the drift velocity of the cells depend on their concentration and on the concentration of the secreted chemical, the original Keller-Segel model takes the form of a generalized mean field Fokker-Planck equation.

The statistical characteristics of trajectories of motile {\it D.d.} cells have been the subject of several recent studies. These include experiments to characterize chemotactic cell movement in homogenous and inhomogenous chemical cues \cite{Amselem-2012,Amselem-2012-PRL,Theves:2009, Song-2006, Andrew-2007,Haastert-2009,Haastert-2010,Flyvbjerg-2011}, and parallel theoretical modeling to reproduce statistical features of the experimental observations \cite{Amselem-2012,Holschneider:2014,Song-2006,Haastert-2010,Flyvbjerg-2011}. Recently, Li {\it et al.} have presented an experimental study of the individual cells in a homogeneous medium \cite{Flyvbjerg-2011}. They have proposed a generalized Langevin equation for the velocity of individuals. Their data-driven modeling showed a "programmed" periodic motion around a persistent direction of motion on short time scales and ordinary diffusive behavior on long time scales. Moreover, it is also well known that a cAMP gradient induces a bias of the position where pseudopodia emerge~\cite{Haastert-2009}. The measured probabilities of pseudopod directions were used to obtain an analytical model for chemotaxis of cell populations~\cite{Haastert-2010}. The prediction of the model are similar to measured chemotactic index of wild-type cells as well as the mutants.  Besides, although it is well-known that the directed movement of the {\it D.d.} cells in response to the chemoattractant cAMP depends both on the absolute value of the local concentration (chemokinesis) and its gradient (chemotaxis), the exact dependency is not well understood.

In this study, we aim to extract the concentration dependencies of the diffusion and drift coefficients in the Fokker-Planck equation (with respect to cell density), by analyzing the experimental trajectories of motile {\it D.d.} cells in Ref.~\cite{Amselem-2012,Amselem-2012-PRL,Theves:2009}. We assume that these coefficients depend on both the local cAMP concentration (the so-called midpoint concentration) and its gradient. The experiments are performed in a microfluidic device (see Section II-B) that generate linear stable gradients between the two inlet concentrations $C_\text{max}$ and $C_\text{min}$. As the cells crawl up the gradient, the average background concentration they experience increases. These experiments systematically explore different steepnesses and cover a wide range of gradients, at which chemotactic behavior is observed. In these experiments, an external flow removes the naturally produced cAMP secreted by the cells to avoid cell-cell signaling. This is completely different from aggregation process where the cell density is much higher and the cells signal each other. We start our analysis by assuming linear dependencies for the diffusion coefficient and the drift velocity of the cells  along the width of microfluidic setup, where a linear gradient is established. We then use the experimental cell trajectories to deduce the coefficients of these linear dependencies at different cAMP gradients. 
\section{II. Experiments}
\subsection{A. Cell Culture}
All experiments were performed by M. Theves~\cite{Amselem-2012,Amselem-2012-PRL,Theves:2009} with {\it Dictyostelium discoideum} AX3 wild type cells. Cells were grown in HL5 medium (7 g/L yeast extract, 14 g/L peptone, 0.5 g/L potassium dihydrogen phosphate, 0.5 g/L disodium hydrogen phosphate, 13.5 g/L glucose, ForMedium Ltd., UK). Cells were starved in shaking suspension of phosphate buffer (pH 6.0, 15 mM KH2 PO4 , 1 mM Na2 HPO4 ) at a density of $2\times10^6$ cells/mL for 5:30 hours. After one hour of starvation, the cells were exposed to periodic pulses of cAMP for the remaining time of starvation. The pulses had a concentration of $50~nM$ and were delivered with a period of 6 minutes.

\subsection{B. Microfluidics} \label{microfluidic}

\begin{figure*}[t]
\includegraphics[width=1.8\columnwidth]{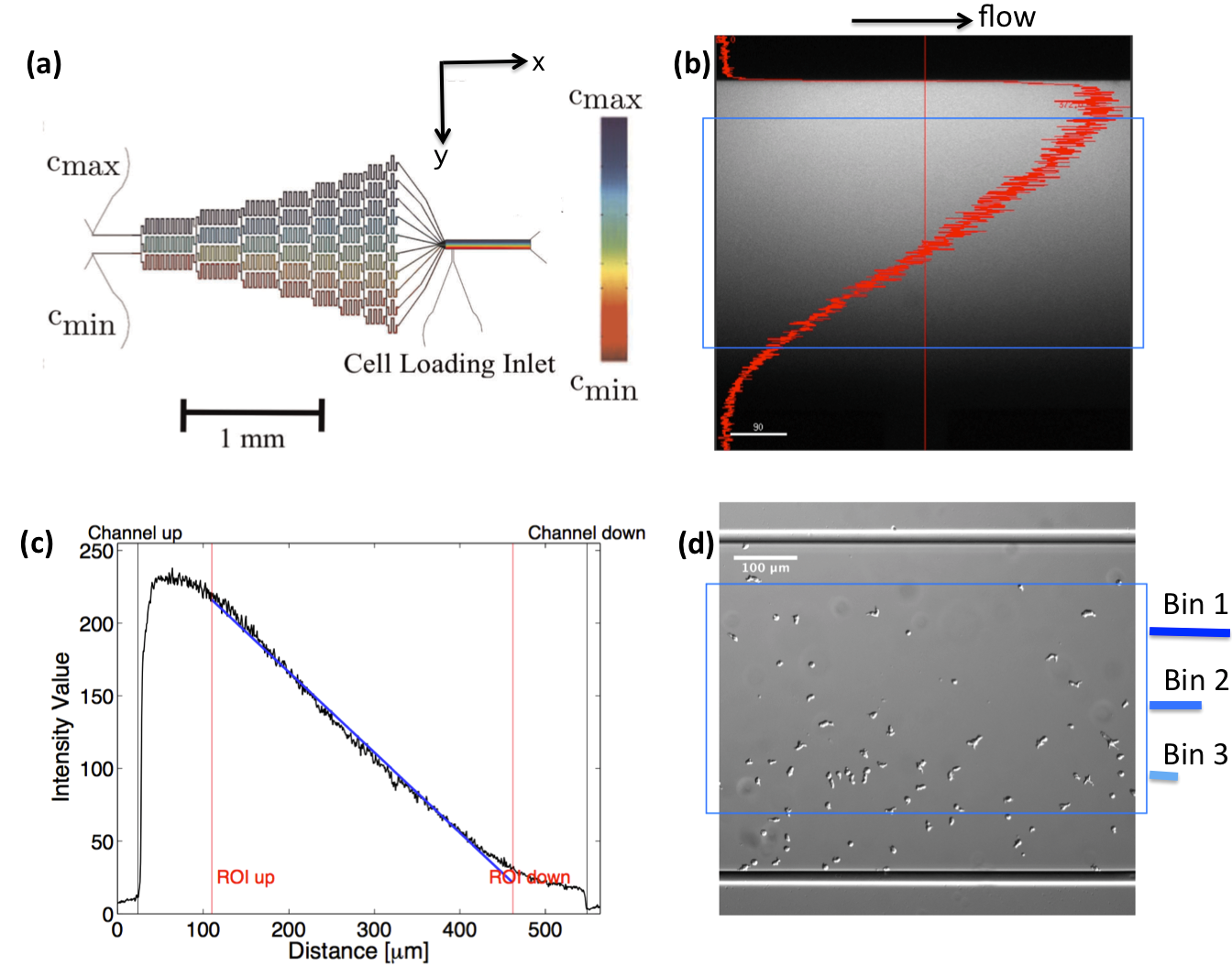}
\caption{ (a) Microfludic Gradient Mixer for Chemotaxis Experiments: two different concentrations flown into the channel inlets undergo steps of diffusive mixing at each branch to form a linear stable gradient in the area of observation. (b,c) Line Profile of fluorescein intensity inside the gradient mixer perpendicular to the flow direction, (d) Differential Interference Contrast (DIC) image, showing the cell population being exposed to the gradient. Only cell trajectories within the region of interest (blue box) are considered for statistics. Moreover, Bin 1 corresponds to the area, with the highest, Bin 3 to the one with the lowest average midpoint concentration experienced by the cells. This figure is used by the permission of M. Theves from his master thesis~\cite{Theves:2009}.}
\label{fig:Setup}
\end{figure*}

A microfluidic gradient mixer~\cite{Song-2006,Whitesides-2000} with given dimensions (width=$525~\mu m$, height= $50~\mu m$) was used to establish a stable linear gradient over a region of $350~\mu m\times 50~\mu m\times3000~\mu m$ in size (see Fig.~\ref{fig:Setup}). The gradients were generated using a pyramidal microfluidic network that provides well-defined concentration profiles with high temporal stability. Throughout the experiment, a constant flow is provided by syringe pumps. The flow provides a constant supply of oxygen and removes all substances released by the cells. This prevents cells from signaling each other, which would perturb the concentration gradient and bias the chemotactic motion. Running at an adjustable average flow velocity of $\bar v = 320~\mu m/s$, the gradient is linear and stable within $d=350$ $\mu m$ in the middle of the channel. Above a lower threshold of $\nabla C_\text{thresh}\sim 10^{-3}~nM/\mu$m cells started to show a directional response. It is important to note, that all gradients have been established by mixing a phosphate buffer solution at one inlet, $C_\text{min} = 0$, together with a solution of cAMP and phosphate buffer $C_\text{max}$ on the opposing inlet. Therefore the gradient
\begin{equation}
\nabla C = C_\text{max}-C_\text{min} = \Delta C/d
\end{equation}
always ranges from zero to this maximum concentration. Due to boundary effects, the profile is distorted near the walls. All cell trajectories within this non-linear area were excluded from statistics.  Moreover, given the dimension of the channel and the dynamics viscosity of the flowing phosphate buffer ($\eta=10^{-3}$ Pa s), one can calculate the shear stress applied on the cells at the imposed mean flow velocity of $\bar v = 320~\mu m/s$ to be $\sigma = 0.038$ Pa.  According to the literature, mechanosensing in {\it D.d.} has been observed above a threshold of $\sigma = 0.5$ Pa~\cite{Mechanotaxis}. We are thus approximately one order of magnitude below the regime where flow induced shear stress would bias the motion of chemotactic cells.
\subsection{C. Cell Tracking}
Differential Interference Contrast (DIC) images were recorded for 180 min, with time resolution of 10 sec and spatial resolution of 1024x1024 pixel (1 pixel=$0.6409~\mu m$), and processed  using Mathworks MATLAB 7.5 with the Image Processing Toolbox~\cite{Amselem-2012,Amselem-2012-PRL,Theves:2009}. All the image processing steps are done by M. Theves {\it et al.} The images were binarized  to distinguish the cells from the background and possible optical artifacts. The cell centroids in each binarized frame were identified. To produce cell trajectories one had to link these locations together in time and space. To achieve this, a customized version of the MATLAB cell tracking algorithm written by Crocker and Grier~\cite{Crocker:1996} was used. This tracking process is consisted of calculating and minimizing
the sum over the squared displacements of all possible links
between the cell positions in two subsequent frames. Interestingly, an analysis of the broken trajectories have shown that more than 90 percent of the cells were lost due to a sudden jump in the cell location or because two cells ran into each other and their center of mass in the binarized image became indistinguishable. In this case, the tracks will end and new ones will start, once the cells separate again. Tracks may
also end when the segmentation program loses a cell due to image
quality problems. Once the cell is detected again, a new track will start.  These different scenarios result in a distribution of tracks of different length with most of them
shorter than the total measurement time. They also result that the number of trajectories (e.g. 582 trajectories in Fig.~\ref{fig:track50nM}) is much greater than the number of cells ($\sim$ 40 cells at $t=0$) in the experiment. Note that during an experimental recording, the number of cells is not constant with time as most of the fast cells move out of the region of interest or new cells enter the field of view. Finally, it is important to note that since the cells begin responding to the cAMP  at different time points, or as the cells collide and new tracks start, the starting time-points of all trajectories are set to zero.

\subsection{D. Selection of Trajectories} 
\label{Selection}
In our analysis, to have a reliable statistics, we keep the number of trajectories during the averaging process constant. Trajectories are selected based on two criteria: (i) they should persist at least 20 min and (ii) within this time interval, the cells should migrate more than 20 $\mu$m in $-\hat y$ direction. The minimum displacement of $20~\mu$m in the direction of gradient, for $t = 20$ min, gives an average motility of $\bar v_y > 1~\mu$m/min. Cells with $\bar v_y < 1~\mu$m/min  are neglected to exclude dead or immobile cells from statistics. As previously mentioned, we lose track of the cells once they collide. Therefore, it is important to note that based on this criteria, if a cell collides with another cell and the time interval between two successive collisions is less than 20 min, this trajectory is excluded from statistics although the cell was crawling with $\bar v_y > 1~\mu$m/min. Eventually, to improve our statistics, long trajectories, are truncated at 20, 40, 60,... min and, if the conditions above are satisfied, trajectories between 20 to 40 min, 40 to 60 min,etc. are considered as new trajectories and the starting time point of each trajectory is set to zero (see Fig.~\ref{fig:track50nM}). 
\section{III. Model}
Nonlinear mean field Fokker-Planck equations can find important applications
in the context of chemotaxis~\cite{Murray-1991}. Here, we attempt to implement an advection-diffusion approach to describe the chemotactic movement of the {\it D.d.} cells experiencing a linear stationary gradient~\cite{Amselem-2012,Amselem-2012-PRL}. The statistical properties of the system are characterized by the values of the model parameters returned after fitting the model to the experimental trajectories. 
\begin{figure*}[t]
\begin{center}
\includegraphics[width=1.8\columnwidth]{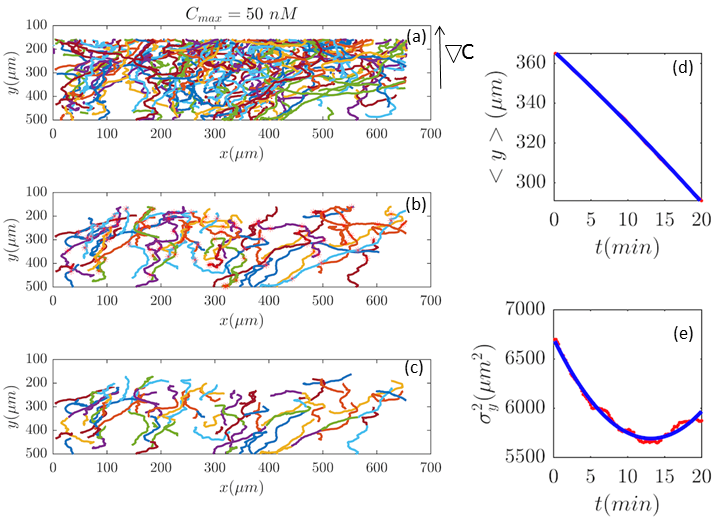}
\caption{(Color online) (a) 582 trajectories tracked in a microfluidic channel with $C_{\text{max}} = 50 \, nM$ ($\nabla C=0.14~nM/\mu m$). Cells migrate on average upwards from the bottom of the channel to the top areas with higher cAMP concentration. (b) 88 trajectories selected (out of 582) based on the two conditions explained in Section II.D. The stars mark the cell positions exactly at $20$ min and (if the trajectory is long enough) at $40$ min, $60$ min and etc. (c) The same trajectories in (b) truncated at $20$ min to keep the number of cells during the averaging process constant. For long trajectories, if the two conditions are satisfied, the tracks between $20$ to $40$ min or from $40$ to $60$ min, etc., are considered as new independent trajectories to improve the statistics. (d,e)  The comparison between experimental data (red lines) and the fitted model (blue line) for $\langle y\rangle$ and $\sigma_y^2$.}
\label{fig:track50nM}
\end{center}
\end{figure*}

To study the chemotactic movement of the {\it D.d.} cells, we consider an advection-diffusion model in which the centroid of the cell's perimeter is represented as the position of a particle. We define an orthonormal basis with the unit vectors $\hat{x}$ and $\hat{y}$, where $\hat{x}$ is the flow direction and $-\hat{y}$ is the direction of the spatial gradient of cAMP (see Fig. \ref{fig:track50nM}).  The position of each cell is given by $\vec{r} = x \hat{x} + y \hat{y}$.  The concentration of the {\it D.d} cells is low enough, so  we can assume that each cell does not sense the presence of the other cells. As stated in Experiment section, microfluidic gradient along $y$ direction is generated by a continuous flow along $x$. To avoid mixing up the issues of chemotaxis in response to the chemoattractant and mechanotaxis under the influence of the shear stress due to viscous forces, we limit our model to the chemotactic movement of the cells along $y$. 
Let us assume that $p(x,y,t)$ denotes the number density of cells at position $(x,y)$ at time $t$. Then, we have 
the probability density $P(y,t)$, which is the original density $p(x,y,t)$ integrated over $x$. The current density along $y$, $J$, reads as
\begin{equation}
 J= -D \p_y P +v P.
\end{equation} 
where $v$ and $D$ are drift velocity and diffusion coefficient, respectively, and $\p_y $ means differentiation with respect to $y$.
Now, the continuity equation for $P$ and $J$ then reads as
\begin{equation}
\p_tP=- \p_yJ,
\end{equation}
where $\p_t $ denotes differentiation with respect to $t$. Using Eqs. (2) and (3), one can find the diffusion-advection equation for the problem as
\bea
\p_t P = \p_y \left( D \p_y P \right) - \p_y (v P). 
\eea
The chemotactic motion of the cells depends on both the absolute local concentration (chemo-kinesis) and its gradients (chemotaxis) \cite{Amselem-2012,Song-2006}. Based on the experiments of Ref. \cite{Amselem-2012}, here we consider a constant spatial gradient of cAMP in the direction of $-\hat{y}$. Therefore, one can expect that both the diffusion coefficient, $D$, and the drift velocity, $v$, depend on the $y$ component of the position vector. Since there is no direct experimental method to determine this dependency, it is plausible to expand the mentioned coefficients in terms of $y$ as
\bea
v = v_0+v_1y+\cdots \label{eq:v}
\eea
and
\bea
D=D_{0}+D_{1}y+\cdots 
 \label{eq:D}
\eea
 We keep the terms only up to the first order of $y$, and treat $v_1$,
 and $D_{1}$ as perturbation coefficients. Here, we assume that the current in the $y$ direction does not depend on $x$. Using equations (4) to (6), one finds

\begin{eqnarray} 
\p_t P=(D_0+D_1 y) \p_y^2 P+ (D_1-v_0-v_1 y) \p_yP - v_1 P. \nonumber \\
\label{eq:master2}
\end{eqnarray}

The mean value of $y$-component of the cells' positions is obtained by
\begin{equation}
\langle y(t) \rangle = \int y \, P(y,t) dy. 
\end{equation}
Differentiating the above expression with respect to time results 
\begin{eqnarray}
\frac{d }{d t}\langle y(t) \rangle = \int dy\, y\,\p_t P(y,t).
\end{eqnarray}

After substituting $ \p_t P(y,t)$ from Eq. (\ref{eq:master2}) and integrating, one can find 
\bea
\frac{d}{d t} \langle y(t) \rangle =v_0+ D_{1}+v_1  \langle y(t) \rangle. \label{eq:dydt} 
\eea
By solving this simple ordinary differential equation we find
\bea
 \langle y(t) \rangle =e^{v_1 t} \left[ \frac{v_0+D_{1}}{v_1}+ \langle y \rangle_0 \right] -\frac{v_0+D_{1}}{v_1},
\eea
where $\la y \ra_0 \equiv \la y \ra |_{t=0}$ denotes the mean initial $y$-position of the cells. As it has been mentioned above, $v_1$ and $D_1$ are the small parameters and in our model, they have been considered as perturbation parameters. After expanding the exponential factor and keeping the terms up to the first order of perturbation parameters, $v_1$ and $D_1$, one can find 
\bea
\langle y \rangle (t) = \langle y \rangle_0 + \left(v_0+ v_1 \langle y \rangle_0+D_{1} \right) t + \frac{1}{2} v_0 v_1 t^2. \label{eq:y_t}
\eea
It is worth mentioning that since terms like $v_1 D_1$ are the second order of perturbation parameters, these terms are dropped. 

The variance of the cells' positions along $y$ is defined as 
$\sigma_y^2(t) \equiv \la y(t)^2 \ra - \la y(t) \ra^2$. Using similar method (see Appendix I for details), one can find $\sigma_y^2(t)$ as 
\bea
\sigma_y^2(t) &=& \sigma^2_y(0)+2 \left[ \sigma^2_y(0)\,v_1+D_{0}+D_{1}\langle y\rangle_0 \right] t \nonumber \\
&+&(2D_{0}v_1+D_{1}v_0)\,t^2, \label{eq:var_y}
\eea
where 
 $\sigma_y^2(0)$  is the initial variance of the cells' positions along $y$. 
 We note that in Eq. (\ref{eq:var_y}), we have kept the terms up to the first order of perturbation parameters as well. 
\section{IV. Results}
Now we are in a position to determine the perturbation parameters of our model based on the experimental trajectories. The mean displacement of chemotactic cells and the corresponding variance can be calculated from the experimental trajectories as defined in Appendix I (Eqs.~\ref{eqn:MeanX}--\ref{eqn:VarianceY}). To characterize the chemotactic behavior of {\it D.d.} cells, based on Eqs. (\ref{eq:y_t}) and (\ref{eq:var_y}), we need to determine the values of $v_0$, $v_1$, $D_0$ and $D_1$. We treat these factors as tuning parameters and find their values simultaneously by fitting (MATLAB, R2016b) the model to the experimental values of $\la y \ra_{\text{exp}}$ and $\sigma_{y,\text{exp}}^2$. Tables \ref{table:parameters}-\ref{table:parameters3} include the best fit values of the above mentioned parameters at different cAMP gradients. In Table \ref{table:parameters}, $\langle y \rangle_0$ and $\sigma^2_y(0)$ denote the fitted mean and standard deviations at time zero.
\begin{figure*}[t]
\begin{center}
\includegraphics[width=0.96\columnwidth]{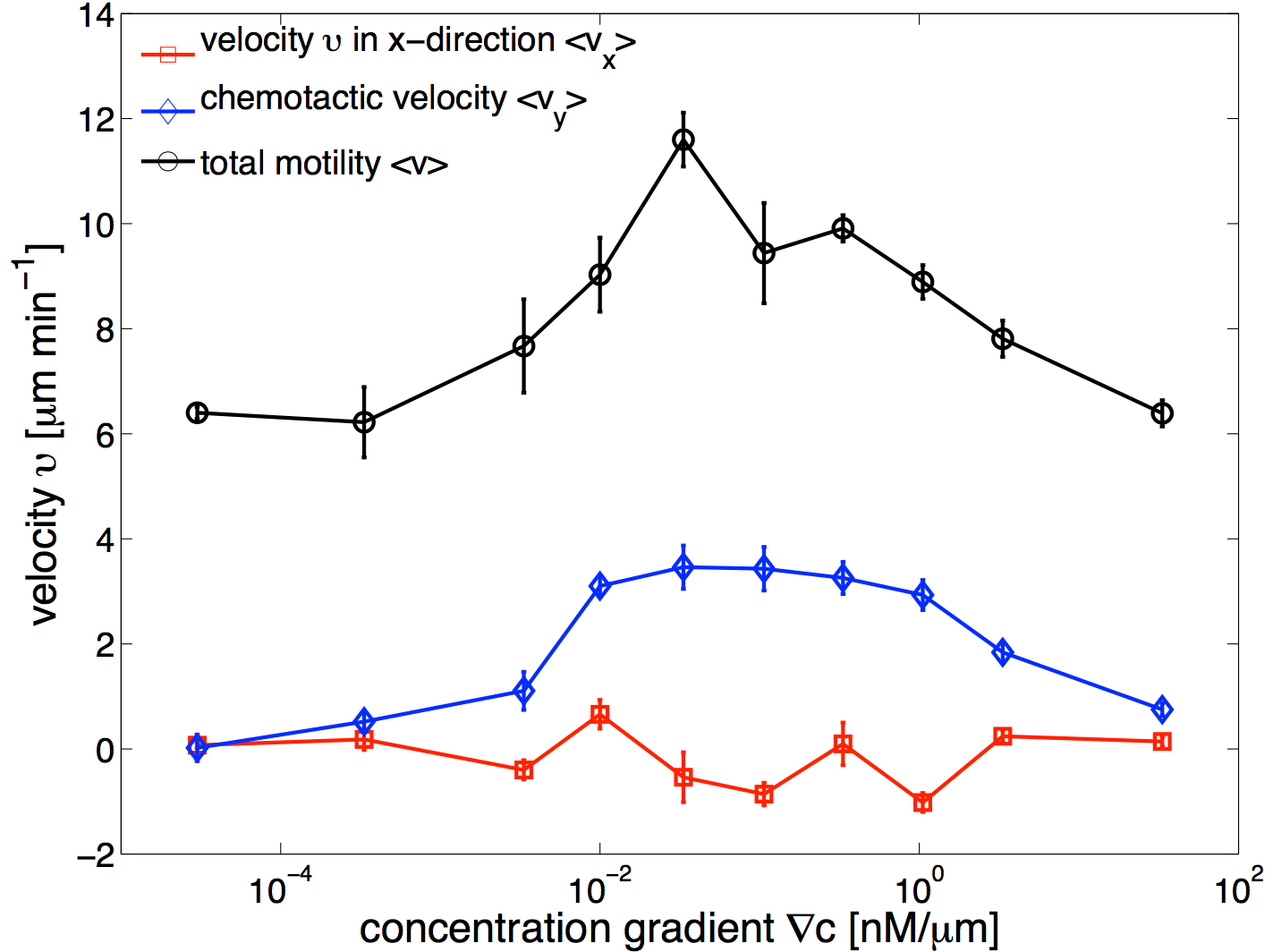}
\includegraphics[width=0.96\columnwidth]{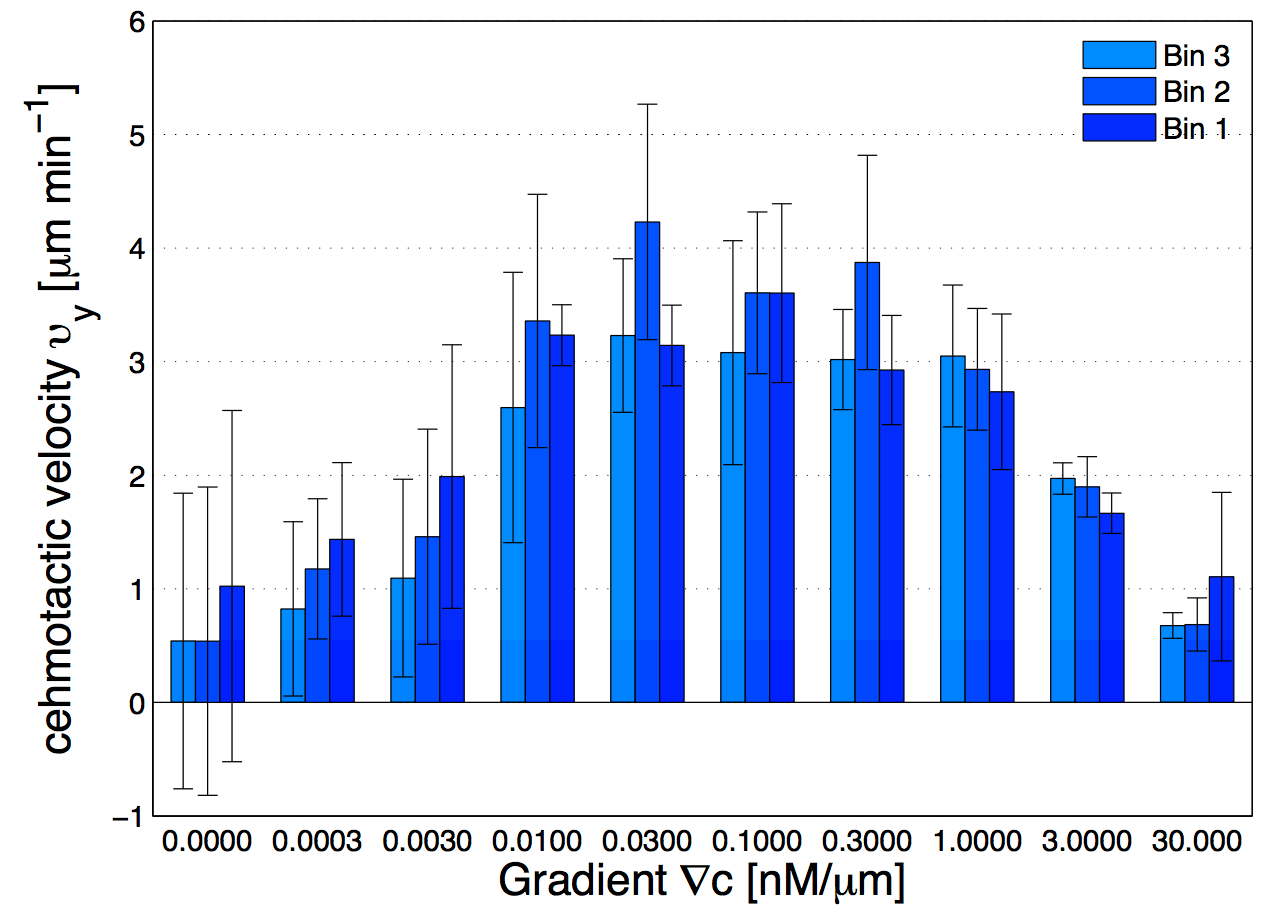}
\caption{(Color online) Independent data analysis done by M. Theves~\cite{Theves:2009} showing (left) chemotaxis as a function of gradient steepness $\nabla C$: above a threshold at $\nabla C \simeq 10^{-3}~nM/\mu m$ cells show a positive (in our coordinate system negative) average velocity in gradient direction ($v_y$) as well as an increased total motility $v$, while the perpendicular velocity component in flow direction ($v_x$) remains random and averages to zero within error bars. For gradients ranging over two orders of magnitude, $10^{-2} nM/\mu m \le\nabla C \le 1~nM/ \mu m$, the chemotactic speed is constant. (right) Average chemotactic velocity $v_y$ as a function of gradient steepness evaluated separately for three different areas, subdividing the region of interest (see Fig.~\ref{fig:Setup}d). The midpoint concentration decreases from bin 1 to bin 3. For shallow gradients, the chemotactic velocity increases with a raise in the average midpoint concentration. This effect seems to reverse for steep gradients above $1~nM/\mu m$, where $v_y$ decreases slightly in higher concentration backgrounds. The figures are used by the courtesy of M. Theves from his master thesis~\cite{Theves:2009}.}
\label{fig:Velocities}
\end{center}
\end{figure*}

Fig.~\ref{fig:track50nM} and Figs.~\ref{fig:track1nM}--\ref{fig:track10000nM} (see Appendix III)  show the comparison between the model and the  experiments at different cAMP concentrations.  The initial number of trajectories before the selection procedure are presented in part (a) of each figure. Trajectories for our analysis are then selected and truncated based on the criteria explained in Section II.D. Selected and truncated trajectories are shown in parts (b) and (c) of each figure, respectively. The initial number of trajectories as well as the number of selected trajectories are different for different cAMP gradients. In parts (d-e) of each figure, the red lines correspond to the experimental data and the blue lines correspond to the results of our model using the fitting parameters of Tables \ref{table:parameters}-\ref{table:parameters3}. The important features of the figures and the tables are summarized below:
\begin{itemize}
\item{The mean position of the chemotactic {\it D.d.} cells, $\langle y \rangle$, decreases almost linearly in time, which shows that the chemotactic cells migrate towards areas with higher cAMP concentration (top areas of the channel). The nonlinear term $v_0 v_1/2$ in Eq.~\ref{eq:y_t} is small for all concentrations (see the last column of Table.~\ref{table:parameters1})}.

\begin{figure*}[t]
\begin{center}
\includegraphics[width=1.7\columnwidth]{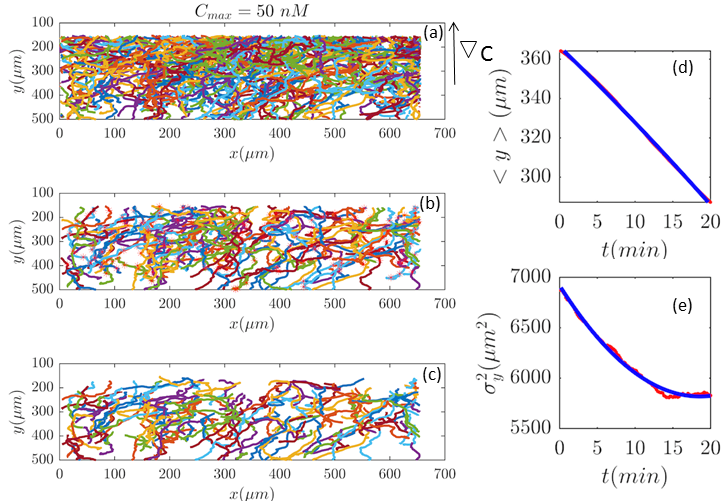}
\caption{(Color online) (a) Trajectories of three different experiments recorded at the same cAMP gradient of $0.14~nM/\mu m$ are combined to improve the statistics. (b) 207 (out of 1097) trajectories, are selected and (c) truncated based on the criteria in Section II.D. These trajectories participate in our analysis which is more than two times the number of selected trajectories in Fig.~\ref{fig:track50nM}. (d,e) The comparison between experimental data (red lines) and the fitted model (blue line) for $ \la y \ra$ and $\sigma_y^2$.}
\label{fig:track50nM-3ExpMixed}
\end{center}
\end{figure*}

\item{The mean square displacement function $\sigma^2_y(t)$ shows decreasing behavior at $C_{max}=10$, $50$, $316$ and $10000~nM$. This trend is an experimental observation, independent of the introduced model, and  has to do with the fact that the cells tend to migrate to areas with high cAMP concentration (top part of the channel). Since in these areas mid-point cAMP concentration is high and most of the cAMP receptors are saturated, therefore the cells slow down and accumulate at the top of the channel. This ``accumulation'' can give rise to a decreasing  $\sigma^2_y(t)$. In the other word, the possible decrease in the variance is due to a drift towards the top areas of the channel.}

\item{The diffusion coefficient in $y$ direction $D_0+D_1\langle y\rangle$ is initially positive for all concentrations but as the cells migrate upwards and the value of $\langle y\rangle$ decreases, it becomes negative for all concentrations except for $C_{max}=1~nM$. We think that this negative diffusion coefficient extracted from the data is an artifact of  the perturbative approximation. To be more exact, the diffusion coefficient depends on the position,
and we have Taylor-expanded it and kept only the zeroth and 
the first order terms (the latter as a perturbative term). While 
the full position-dependent diffusion coefficient should probably
be non-negative, there is no such restriction on its truncated form
(which contains only the first two terms): it is just a parameter
which is determined through a best fit. Of course the parameters
should respect the non-negativity of the variance, and they do, 
as it is seen that the fitted variance does not become negative.
 }

\item{We define the mean drift velocity of chemotactic {\it D.d.} cells as
\begin{equation}
v_\text{drift}= \frac{d\langle y \rangle}{dt } = v_0 + v_1 \langle y \rangle_0 + D_1 + v_0 v_1 t,
\end{equation}
which shows that the drift velocity depends not only on $v_0$ and $v_1$ but also on $D_1$. The coefficient $v_0v_1$ is a small number for different cAMP gradients (see Table.~\ref{table:parameters1}), therefore $v_\text{drift}$ is essentially constant in time. The extracted values of drift velocity at time zero are listed in the fifth column of Table.~\ref{table:parameters1}. It is interesting that the drift velocity in $y$ direction doesn't depend significantly on the cAMP gradient and fluctuates around 4 $\mu m$/min. This is consistent with an independent data analysis performed by M. Theves~\cite{Theves:2009} (see Fig.~\ref{fig:Velocities}): within a plateau, ranging from $10^{-2}~nM/\mu m\le \nabla C \le 1~nM/\mu m$ over two orders of magnitude, the chemotactic velocity is almost constant. For gradients above $\nabla C = 1~nM/\mu m$ the directionality of movement is decreased. Exceeding an upper threshold of $\nabla C_{up}\sim 10^{2}~nM/\mu m$, the cell motion becomes isotropic again.
}

\item{For all gradients, while the cells crawl up the gradient, the magnitude of $\langle v_y\rangle=v_0+v_1\langle y\rangle$ decreases as the midpoint concentration increases. However, the independent data analysis of M. Theves~\cite{Theves:2009} shows a transition: for shallow gradients, right after the onset of chemotaxis $\nabla C= 0.003~nM/\mu m$, $v_y$ increases as the background concentration rises. This effect reverses for steep gradients $\nabla C> 0.3~nM/\mu m$ (see right panel of Fig.~\ref{fig:Velocities}).}
\end{itemize}
\begin{table*}[t]
\caption{The mean initial positions of the cells and the corresponding variances at time zero for different cAMP concentrations.}
\begin{center}
\begin{tabular}{cccccc}
 $C_{\text{max}}( nM)$ &$\nabla C~(nM/\mu m)$& $ \langle y \rangle_0 \, (\mu m)$&$\sigma_x^2(0)(\mu m ^2)$&$\sigma_y^2(0)(\mu m ^2)$ \\
\hline $1$ &$0.003$& $373.39$&$10880$&$5454.2$  \\
$10$ &$0.03$ & $347.22$&$25893$&$6857.1$  \\
$31.6$ &$0.09$ & $417.88$&$24136$&$7608.9$ \\
$50$ &$0.14$ &$365.99$&$27253$&$6707.3$  \\
$100$ &$0.29$ &$410.24$&$32453$&$4719.2$  \\
$316$ &$0.9$ &$371.13$&$30854$&$7033.1$  \\
$10000$ &$28.6$ &$382.33$&$27665$&$6522.6$  \\
\hline
\end{tabular}
\end{center}
\label{table:parameters} 
\end{table*}
\begin{table*}[t]
\caption{Drift coefficients for different cAMP gradients. The mean drift velocity of the cells at time zero is presented in the fifth column.}
\begin{center}
\begin{tabular}{cccccc}
 $C_{\text{max}}( nM)$ & $v_0 (\mu m / min)$ & $v_1 (1/min)$ & ~$v_0+v_1 \langle y \rangle_0  $& ~$v_0+v_1 \langle y \rangle_0+D_1 $& ~$v_0v_1/2$  \\
\hline $1$ &  $2.87$ &$-0.017$ & $-3.32$ & $-2.94$&$-0.024$ \\
$10$  &$-7.62$ &$-0.012$ & $-11.69$ & $-4.06$&$0.045$ \\
$31.6$  & $-7.27$ &$-0.010$ & $-11.04$ & $-4.66$&$0.033$ \\
$50$  &$2.04$ &$-0.016$ & $-3.95$ & $-3.45$&$-0.017$ \\
$100$  &$3.36$ &$-0.016$ & $-3.16$ & $-3.08$&$-0.027$ \\
$316$  &$-8.98$ & $-0.013$ & $-13.77$ & $-4.83$&$0.058$ \\
$10000$ &$-10.71$ & $-0.015$ & $-16.36$ & $-3.97$&$0.079$ \\
\hline
\end{tabular}
\end{center}
\label{table:parameters1} 
\end{table*}

\begin{table*}[htbp]
\caption{Diffusion coefficients in $y$ direction for different cAMP gradients.}
\begin{center}
\begin{tabular}{cccccc}
 $C_{\text{max}}( nM)$& $D_0$ & $D_1 $& $D_0+D_1 \langle y \rangle_0$ & ~~$2D_0v_1+D_1v_0$ ~ & ~$2(\sigma_y(0)^2v_1+D_0 + D_1 \langle y \rangle_0)$   \\
\hline  $1$ &$-49.10$ & $0.37$ & $89.50$&$2.69$&$-1.81$ \\
$10$ &$-2612.60$ &$7.63$ & $38.23$&$3.09$&$-84.38$\\
$31.6$ & $-2581.40$ &$6.38$ & $86.67$&$0.27$&$35.76$ \\
$50$ &$-148.60$ &$0.50$ & $32.52$&$5.87$&$-154.38$\\
$100$  &$129.80$ &$0.09$ & $164.54$&$-3.84$&$179.00$ \\
$316$  &$-3270.57$ & $8.90$ & $45.22$&$4.15$&$-91.00$\\
$10000$&$-4690.80$  &$12.39$ & $46.59$ &$6.00$&$-99.65$\\
\hline
\end{tabular}
\end{center}
\label{table:parameters3} 
\end{table*}
\section{V. Discussion} \label{Discussion}
We have analyzed large data sets of {\it D.d.} chemotaxis in
linear gradients of cAMP recorded by Theves {\it et al.} in a microfluidic setup~\cite{Amselem-2012,Amselem-2012-PRL,Theves:2009}. Data sets with different
steepnesses of cAMP gradient were included in our analysis, covering a large range of gradients, in which chemotactic behavior was observed. Inspired by the experimental conditions of Ref.~\cite{Amselem-2012}, we introduced a minimal phenomenological model that explicitly incorporated the dependency of diffusion matrix and velocity of the cells on their positions which corresponds to the position dependence of the local concentration of chemotactic cues. Based on this model, we extracted the physical properties of the chemotactic {\it D.d.} cells using the mean and variance of the experimental cell tracks. What is the benefit of this phenomenological model? As highlighted previously, chemotactic movement of the cells depend on both the chemoattractant gradient and the average ambient chemoattractant concentration (midpoint concentration). In the microfluidic setup of Theves {\it et al.}~\cite{Amselem-2012}, the cells are exposed to a constant gradient, while the midpoint concentration increases when the cells are moving up the gradient. Traditionally, chemotactic cell motion is described by Langevin-type equation where for each cell track, the velocity and acceleration of the cells are calculated at each point by finite differences from the cell
positions~\cite{Amselem-2012, Amselem-2012-PRL,Flyvbjerg-2005}. Therefore in these types of analysis midpoint concentration is globally averaged out. Other quantities such as chemotactic index, defined as the distance moved in
gradient direction divided by the total distance, are also averaged quantities where information on mid-point concentration is lost.  
However in our analysis, instead of velocity and acceleration, we work directly with spatial position of the cells  and explicitly include  the dependence of the diffusion coefficient and the drift velocity on the
midpoint concentration. Taylor expansion of these coefficients up to the first order in $y$ leads to a closed set of equations that can be solved to obtain the fitting parameters. It is worth to check the effect of the dependency of the diffusion and velocity on the local concentration. To do this, let us assume that the drift velocity and the diffusion coefficient were constant. We denote the constant drift velocity and diffusion constant by $\tilde{v}_0$ and $\tilde{D}_0$, respectively, to obtain 
\bea
\langle y \rangle (t) &=& \langle y \rangle_0 + \tilde{v}_0 t, \label{eq:y_t2} \\
\sigma_y^2(t) &=& \sigma_y^2(0) + 2 \tilde{D}_0 t. \label{eq:var_y2}
\eea
These equations predict a linear dependency on $t$, in both the mean and the variance of the position, which is not consistent with the experimental data especially in the variance of $\sigma_y^2$ (see Fig.~\ref{fig:track50nM} and Figs.~\ref{fig:track1nM}--\ref{fig:track10000nM}). In fact it has been shown in Ref.~\cite{Khorrami-2012}, that any linear diffusion model (even anomalous), which enjoys both time translational invariance and space translational invariance leads to means and variances which are at most linear in time. This is a motivation to use drift and diffusion parameters which do depend on the position. There is an obvious position-dependence in our system: C (the concentration of cAMP) does depend on the coordinate $y$. Assuming that the drift and diffusion parameters do depend on C, one is left with a $y$-dependence in the drift and diffusion parameters. A simple manageable model is to Taylor-expand this $y$-dependence and keep only terms which are up to first order in $y$. The result is a first order perturbation model, which has been studied here. We emphasize that the experimental conditions of Ref.~\cite{Amselem-2012} fulfill the necessary conditions of the mentioned study.
\begin{figure*}[t]
\begin{center}
\includegraphics[width=1.5\columnwidth]{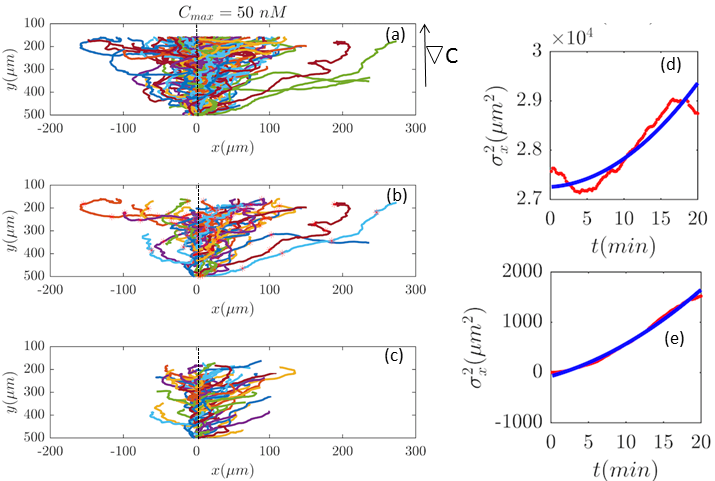}
\caption{(Color online) (a) The same trajectories as in Fig.~\ref{fig:track50nM} which are shifted to $x=0$, (b) selected and (c) truncated based on the criteria in Section II. D. The $y$ dependency of all trajectories are kept as before because the mid-point concentration is different along the width of the channel. (d) and (e) show the behavior of $\sigma_x^2$ as a function of $t$, before and after shifting all the tracks to $x=0$, respectively.  Red lines correspond to the experimental data and blue lines correspond to fitted quadratic polynomials, respectively.}
\label{fig:track50nM-XShift}
\end{center}
\end{figure*}

In previous studies, wild-type and mutated epithelial canine kidney cells, it has been shown that the cell dynamics can be characterized by an anomalous diffusion~\cite{Dieterich-2008}. In particular, mean squared displacement shows a super-diffusive behavior. This super-diffusive behaviour was also observed in the mean square displacement of Hydra cells~\cite{Upadhyaya-2001}. However, experimental trajectories of chemotactic {\it D.d.} cells in Ref.~\cite{Flyvbjerg-2011}, were interpreted by a data-driven model with purely diffusive behavior. As we mentioned above, a pure diffusive model can not explain the non-linear behavior observed in the experimental data of Ref.~\cite{Amselem-2012}. 

In our analysis, we observed that at all concentrations $D_0+D_1\langle y\rangle$ decreases with time and becomes negative for concentrations of $10,~31.6,~316,$ and $10000~nM$. In order to make sure that negative diffusion coefficients are not due to our low statistics after the selection procedure, we combined trajectories of three different experiments performed at the same cAMP gradient, namely $\nabla C=0.14~nM/\mu m~(C_\text{min}=0,~C_\text{max}=50~nM)$. Comparison between Fig.~\ref{fig:track50nM} and Fig.~\ref{fig:track50nM-3ExpMixed} shows a similar decreasing behavior in $\sigma^2_y(t)$. Indeed, an absorbing point on top of the channel can produce a decreasing variance, not through the diffusion but through the upstream drift. Let us suppose that $D_0=D_1=0$. Then, according to Eq. \ref{Eq:sigmaY},  $\sigma^2_y(t)=\sigma^2_y(0) \exp(2 v_1 t)$. If everything is expanded up to first order in $v_1$, then the result is $\sigma^2_y(t)=\sigma^2_y(0) (1+2v_1 t)$. As $v_1\le 0$, it seems that $\sigma_y^2(0)$ could become negative after a while. But that is an artifact of the approximation. We intended to find position-dependence of diffusion and drift coefficients. Since the exact position-dependence is not known, even if  the inhomogeneity of the surface is known, we expanded the diffusion and drift coefficients in power of the position $y$. That the time dependence of the variance does match the experiments, means that the method works. But the perturbative parameters should not be misleading.

Furthermore, with our model, we can test directly the space-time symmetries of the cell movement. Based on the reports of the experiments the gradient in the $y$ direction is homogeneous in $x$. However, the cell tracks shown in the panel (a) of Figs.~\ref{fig:track50nM},~\ref{fig:track1nM}, \ref{fig:track10nM}, \ref{fig:track100nM}, \ref{fig:track316nM}, and \ref{fig:track10000nM} seem to show a drift in positive $x$ direction (in addition to the chemotactic drift in -$y$ direction). To check the spatial homogeneity in the $x$ direction, we shifted all the tracks of  Fig. 2 to $x = 0$ (see Fig. 5). It is interesting that in this case $\sigma_x^2(t)$ is not a pure translation of the same function for unshifted trajectories (see Fig. 5). It seems that the behavior of the function depends on the initial condition. This non-pure shift in $\sigma^2_x(t)$ could be a hallmark of correlation between the displacement of individual cells along $x$ direction and their initial $x$-positions (see Appendix II). Apparently, the system does not have the translational symmetry along the $x$ direction. This is surprising, since analysis in Section II.B. shows that with flow speed of $320~\mu m/s$ we are far below the regime, where mechanotactic effects have been observed in {\it D.d.} cells. However, the authors of Ref.~\cite{Mechanotaxis} conducted their experiments with vegetative cells. This suggests that starvation may increase the mechanosensitivity of {\it D.d.} cells. We emphasize that with this correction all of our analysis in the $y$ direction is still valid, given that the current in the $y$ direction does not depend on $x$. This assumption is nothing but a mean-field approximation.

To improve our statistics, we have divided long mother trajectories to shorter ones and if the criteria in Section. II.D are satisfied, we have included daughter trajectories as completely independent tracks in our analysis. The main difference between these new daughter trajectories is the average midpoint concentration that the cells experience as they crawl up the gradient. This corresponds to moving up  from Bin 3 to Bin 1 in Fig.~\ref{fig:Setup}d,  where in each Bin cells are exposed to a different average midpoint concentration. Detailed analysis by Theves {\it et al.} have shown that (see the right panel of Fig.~\ref{fig:Velocities}) with a raise in the average midpoint concentration the average chemotactic velocity $v_y$ doesn't show any clear trend for intermediate gradients, $10^{-2} nM/\mu m \le\nabla C \le 0.3~nM/ \mu m$. However, for shallow gradients $v_y$ increases with midpoint concentration and for steep gradients it decreases. Most of our analysis are done at intermediate gradients where the variation in chemotactic velocity $v_y$ between three different Bins is less than 25 percent. Exemplary, at $\nabla C=0.3~nM/\mu m$, average chemotactic velocity changes from $\sim 3~\mu m/min$ to $\sim 4~\mu m/min$ for three bins with different midpoint concentrations of $17~nM, 50~nM$ and $83~nM$. At steep gradients larger than 0.3 $nM/\mu m$, the variations in $v_y$ is even less than 10 percent.  Thereby we believe that shorter daughter trajectories which belong to one mother long trajectory, do not significantly differ in their chemotactic properties. In other words, by dividing long mother trajectories to shorter daughter tracks, we don't introduce new types of trajectories with completely different statistical properties.

In the present work, even though we have analyzed a substantial amount of data, much larger data sets with longer trajectories would be required in order to improve our statistics. Possible future experiments with wider microfluidic channels can be helpful to obtain longer trajectories. Experiments with lower cell density (to avoid cell-cell collision) can also help us to obtain longer trajectories, as the cells after collision are indistinguishable from each other and two new trajectories are detected by the cell tracking algorithms.
 
In summary, we have analyzed the experimental data of chemotactic {\it D.d.} cells in the linear gradient of cAMP. In order to have a reliable statistics, we kept the number of trajectories during our analysis constant. Trajectories were selected based on two criteria: (i) they should persist at least 20 min, and (ii) within this time interval, the cells should have migrated more than 20 $\mu$m in the direction of the the gradient of cAMP. We have shown that by introducing an advection-diffusion model that includes the position dependence on the cAMP concentration, a quantitative description of experimental cell tracks of the amoeba {\it D.d.} is achieved. Our analysis goes beyond a pure diffusive model and shows that the super-diffusive behavior can dominate at larger time scales. Specifically, while in a conventional advection-diffusion model both the mean and the variance are linear in time, here in both cases terms arise which are quadratic in time. 

In future study, we aim to apply our analysis to the trajectories of cells migrating on surfaces of differing composition. In a recent study, it has been shown that {\it D.d.} cells migrate similarly on surfaces with various chemical composition~\cite{Losert:2016}. As the substrate composition changes, the cells regulate forces generated by actomyosin network to maintain optimal cell-surface contact area and adhesion. We will assess migration trajectories of the cells on different surfaces and investigate the variations in the fitting parameters of our model. Furthermore, we aim to extend our analysis to the trajectories of mutant cell lines that single or multiple components of the chemotactic signaling  pathway are deficient and consequently the character of the cell trajectories may change considerably. Structural differences between the trajectories of wild-type and mutant cells may reflect important information about the role of the various proteins in the signaling pathway of {\it D.d.} cells, which possibly can not be resolved in the models that mid-point concentration informations are averaged out. The  objective is to correlate various parameters  of  our model to the  key molecular players involved in chemotaxis. This can provide a link between the observed macroscopic dynamics and the underlying microscopic mechanism which is an important goal in the field of eukaryotic chemotaxis.
\section{Acknowledgment}
We are deeply grateful to M. Theves, E. Bodenschatz, G. Amselem and C. Beta for sharing the experimental data of Ref. \cite{Amselem-2012} and A. Bae for critical reading of the manuscript. Z.E. and F.M.-R are grateful to A. Celani, R. Golestanian and L. Mollazadeh-Beidokhti for helpful discussions. Z.E. and F.M.-R. acknowledge the hospitality of ICTP in Trieste, where some parts of this work is done. F.M.-R. acknowledges the hospitality of MPIDS-LFPN Group in G\"{o}ttingen, where this work was initiated. M.K. acknowledges the support of the research council of the Alzahra University. A. G. acknowledges the support of the MaxSynBio Consortium which is jointly funded by the Federal
Ministry of Education and Research of Germany and the Max Planck Society.
\section*{Appendix I}
Here we show how to derive Eq.~(\ref{eq:var_y}). Indeed, for what follows we do not need the exact form of $P(y,t)$ itself, but just the time dependence of its moments.
$\langle y(t)^2 \rangle$ is defined as
\bea
 \langle y(t)^2 \rangle = \int y^2  P(y,t) dy. 
\eea 
We can directly obtain an equation for the time evolution of $ \langle y(t)^2 \rangle $ by multiplying the master equation, Eq.~(\ref{eq:master2}), by $y^2$ and integrate over $y$. This results in
\begin{eqnarray} 
\nonumber\frac{\p }{\p t}\int  dy \, y^2  P(y,t) &=& \int  dy \, y^2 \left[ (D_{0}+D_{1}y)\,\partial_y ^2P \right. \\
&+& \left. (D_{1}-v_0-v_1 y)\partial_yP-v_1P \right]. \label{eq:IntMaster2}
\end{eqnarray}
The left-hand side of this equation is simply equal to $\frac{d  \langle y(t)^2 \rangle }{d t}$. 
We apply partial integration to the right-hand side and obtain
\bea
\frac{d}{d t} \langle y(t)^2 \rangle = 2D_0 + 2(v_0+2 D_{1}) \langle y(t) \rangle + 2 v_1  \langle y(t)^2 \rangle. 
\eea
Since $\sigma_y^2(t) \equiv \la y(t)^2 \ra - \la y(t) \ra^2$, and $ \frac{d}{d t}\sigma_y^2(t) = \frac{d}{d t}\la y(t)^2 \ra -2\la y(t) \ra \frac{d}{d t} \la y(t) \ra$, one finds
\bea
 \frac{d}{d t} \sigma_y^2(t) =2 D_{0}+ 2D_{1}\langle y(t) \rangle+2 v_1 \sigma_y^2(t), \label{eq:app-dif_sigma}
\eea
where according to Eq. (\ref{eq:dydt}), $d \langle y \rangle/dt$ has been replaced by $v_0 + D_1 + v_1 \langle y(t) \rangle$. The solution of Eq. (\ref{eq:app-dif_sigma}) is found as
\bea
  \sigma_y^2(t) &=& \left[ \sigma_y^2(0)+\frac{ D_{0}+ D_{1}\langle y(0) \rangle}{v_1} +\frac{ D_{1}v_0}{2v^2_1} \right] e^{2v_1 t}\nonumber\\
  &-& \frac{ D_{1}v_0}{v_1} t- \left[ \frac{ D_{0}+ D_{1}\langle y(0) \rangle}{v_1} +\frac{ D_{1}v_0}{2v^2_1} \right],\label{Eq:sigmaY}
\eea
where $ \sigma_y^2(0) $ denotes the initial variance of the cells' positions along $y$. After expanding the above equation and keeping the terms up to the first order of $v_1$ and $D_1$, one has
\bea
\sigma_y^2(t) &=& \sigma^2_y(0)+2 \left[ \sigma^2_y(0)\,v_1+D_{0}+D_{1}\langle y\rangle_0 \right] t \nonumber \\
&+&(2D_{0}v_1+D_{1}v_0)\,t^2. \label{eq:app-var_y}
\eea

The mean displacement of chemotactic cells, in both $x$ and $y$ directions, and the corresponding variances can be calculated from the experimental data as follows
\begin{eqnarray}
\langle x \rangle_{\text{exp}}(t)  &=&\frac{1}{N} \sum \limits_{i=1}^{N} x_i(t),\label{eqn:MeanX}\\
\langle y \rangle_{\text{exp}}(t)&=&\frac{1}{N} \sum \limits_{i=1}^{N} y_i(t),\\
\sigma^2_{x,\text{exp}}(t)&=&\frac{1}{N}\sum \limits_{i=1}^{N} \left[ x_i(t)-\langle x \rangle_{\text{exp}}(t) \right]^2,\\
\sigma^2_{y,\text{exp}}(t)&=&\frac{1}{N}\sum \limits_{i=1}^{N} \left[ y_i(t)-\langle y \rangle_{\text{exp}}(t) \right]^2,
\label{eqn:VarianceY}
\end{eqnarray}
where $N$ denotes the number of cells. 
\section*{Appendix II}
Here we show how shifting the cells' tracks to $x=0$ can affect on the variance of the x-component through the time, $\sigma_x^2(t)$. Let ${\displaystyle x_i(t)}$ be the position of the $i$'th particle in $x$-direction at time $t$. The displacement of the $i$'th particle in $x$-direction through the time is ${\displaystyle z_i(t) = x_i(t)-x_i(0)}$. Simply, one has $ \la z(t) \ra = \la x(t) \ra - \la x(0) \ra$ where $\la x(t) \ra$ and $\la x(0) \ra$ denote the mean values of $x$-component of the particles at time $t$ and $t=0$, respectively, and $ \la z(t) \ra$ is the mean displacement of the particles. It is worth mentioning that for example $\la x(t) \ra \equiv 1 / N \sum_i x_i(t)$, where $N$ denotes the number of cells.
In order to find the variance, first we note that 
\bea
x_i(t)-\la x (t) \ra =[z_i(t)-\la z (t) \ra ]+[x_i(0)- \la x(0) \ra].
\eea
After squaring both sides of the above equation and averaging, one finds
\bea
\left \la \left[x(t) - \la x(t) \ra \right]^2 \right \ra &=& 
 \left \la \left[z(t) - \la z(t) \ra \right]^2 \right \ra
+ \left \la \left[x(0) - \la x(0) \ra \right]^2 \right \ra \nonumber\\ 
 &+& 2 \left\la \left[ x(0) - \la x(0) \ra \right] \left[ z(t) - \la z(t) \ra \right] \right\ra.  \nonumber \\ \label{A3E1}
 \eea
The covariance of $z(t)$ and $x(0)$ is defined as $\mathrm{cov}[z(t),x(0)] \equiv \left\la (z(t) - \la z(t) \ra) (x(0) - \la x(0) \ra) \right\ra$. This quantity provides a measure for the strength of the correlation between two stochastic variables. 
Using the definition of the variance and covariance, Eq. (\ref{A3E1}) can be written as
\bea
\sigma^2_x(t)=\sigma^2_z(t)+\sigma^2_x(0)+2 \, \mathrm{cov}[z(t),x(0)].
\eea
We see that when $z(t)$ and $x(0)$ are independent, one has $\la z(t) x(0) \ra = \la z(t) \ra \, \la x(0) \ra$ and $\mathrm{cov}[z(t),x(0)] $ becomes zero. In this case, ${ \sigma^2_x(t)}$ differs from ${\displaystyle \sigma^2_z(t)}$ by just a constant shift.
In other words, the necessary and sufficient condition for a pure shift
in the variances of $\sigma_x^2(t)$ and $\sigma_z^2(t)$ is vanishing of the covariance of $z(t)$ and $x(0)$.

\section*{Appendix III}

As we discussed in the main text, the experiments had been done in different cAMP concentrations. Here we present the trajectories, before and after selection procedure, as well as the corresponding analysis for the cAMP concentrations of $C_{\rm max} = 1, \; 10, ~ 31.6, ~ 100, ~ 316,$ and $10000$ nM.

\begin{figure*}[t]
\begin{center}
\includegraphics[width=1.5\columnwidth]{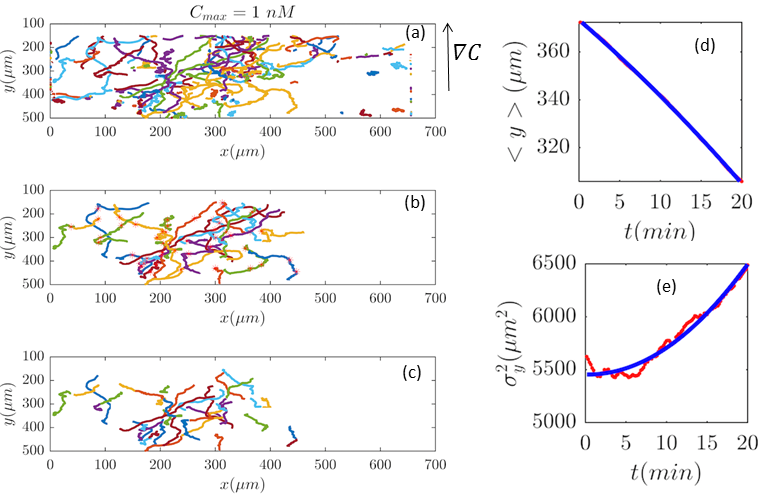}
\caption{(Color online) In an experiment with $C_{max}=1~nM~(\nabla C=0.003~nM/\mu m)$, out of 282 trajectories shown in panel (a), after applying the selection conditions of Section II.D, only 41 trajectories are selected in panel (b) and truncated in panel (c). The comparisons between the experimental data (red) and the model (blue) are presented in panels (d) and (e).}
\label{fig:track1nM}
\end{center}
\end{figure*}

\begin{figure*}[htbp]
\begin{center}
\includegraphics[width=1.5\columnwidth]{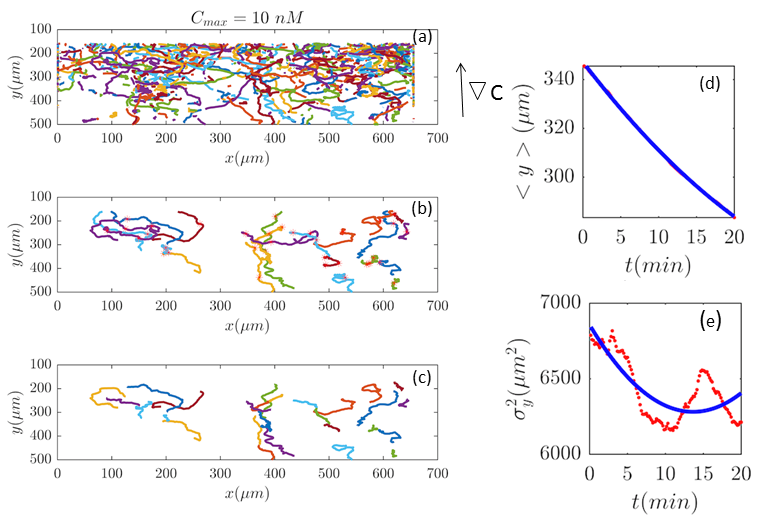}
\caption{(Color online) (a-c) Based on our selection criteria in Section.II.D, only 27 (out of 815) trajectories participate in our analysis for $C_{max}=10~nM~(\nabla C=0.028~nM/\mu m)$. (d) and (e) show $ \langle y \rangle$ and $\sigma^2_y$ plotted for both experimental data (red) and the model (blue).}
\label{fig:track10nM}
\end{center}
\end{figure*}
\begin{figure*}[htbp]
\begin{center}
\includegraphics[width=1.5\columnwidth]{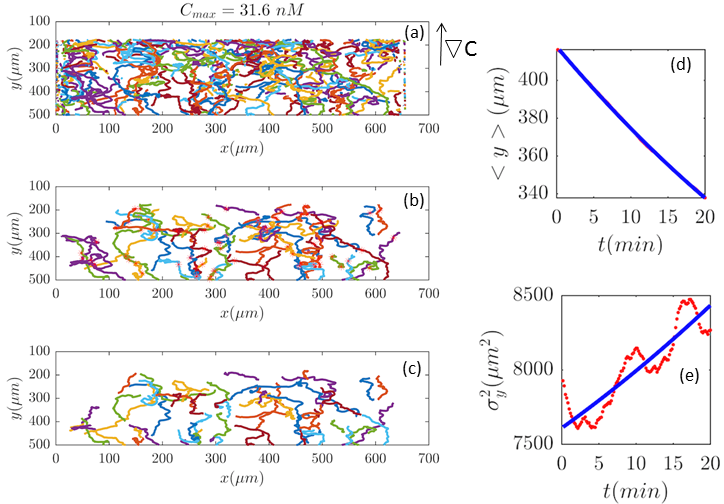}
\caption{(Color online) (a-c) 61 (out of 650) trajectories satisfy the selection conditions of Section II.D for gradient of $\nabla C=0.09~nM/\mu m$ ($C_{max}=31.6~nM$). In panels (d) and (e) the outcome of experimental data and the model are compared.}
\label{fig:track31_6nM}
\end{center}
\end{figure*}
\begin{figure*}[htbp]
\begin{center}
\includegraphics[width=1.5\columnwidth]{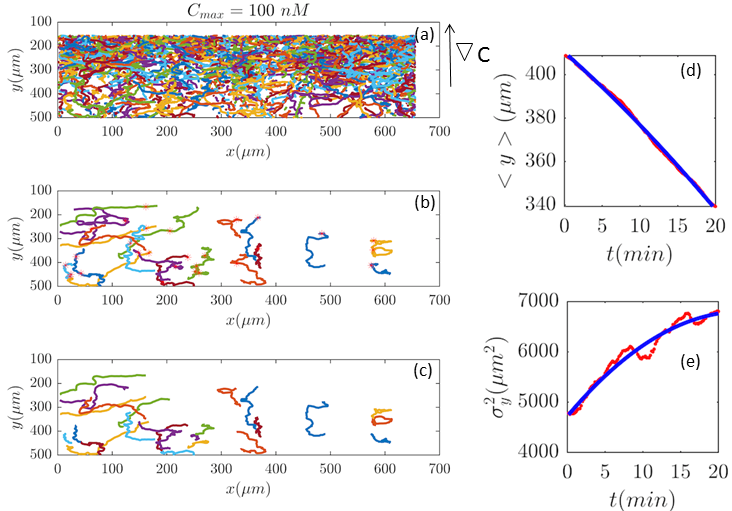}
\caption{(Color online) Out of 2321 trajectories in panel (a), only 25 are selected in panel (b) and truncated in panel (c)  for $C_{max}=100~nM~(\nabla C=0.28~nM/\mu m)$. A large number of trajectories, either show immobile cells or become discontinuous as the cells collide. Again panels (d) and (c) are the comparison between the data and the model.}
\label{fig:track100nM}
\end{center}
\end{figure*}
\begin{figure*}[htbp]
\begin{center}
\includegraphics[width=1.5\columnwidth]{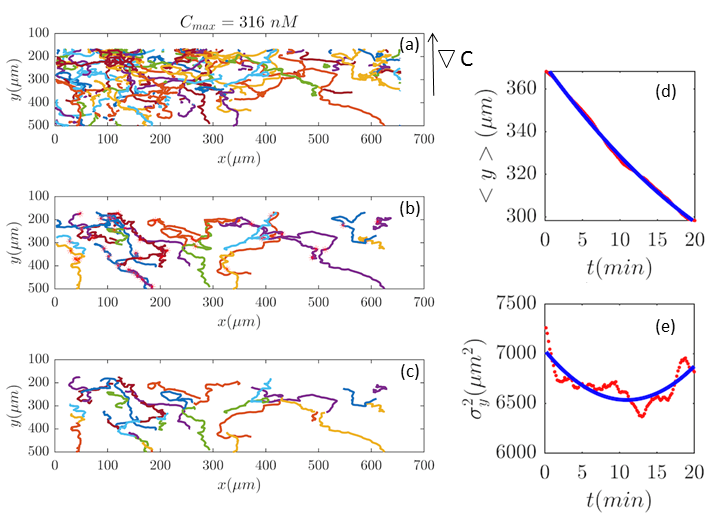}
\caption{(Color online) (a-c) 39 trajectories (out of 254) are participating in our analysis for $C_{max}=316~nM~(\nabla C=0.9~nM/\mu m)$. $y$ and $\sigma^2_y$ are calculated from the truncated trajectories (red lines) and compared with the fitted model (blue lines) in panels (d) and (e).}
\label{fig:track316nM}
\end{center}
\end{figure*}
\begin{figure*}[htbp]
\begin{center}
\includegraphics[width=1.5\columnwidth]{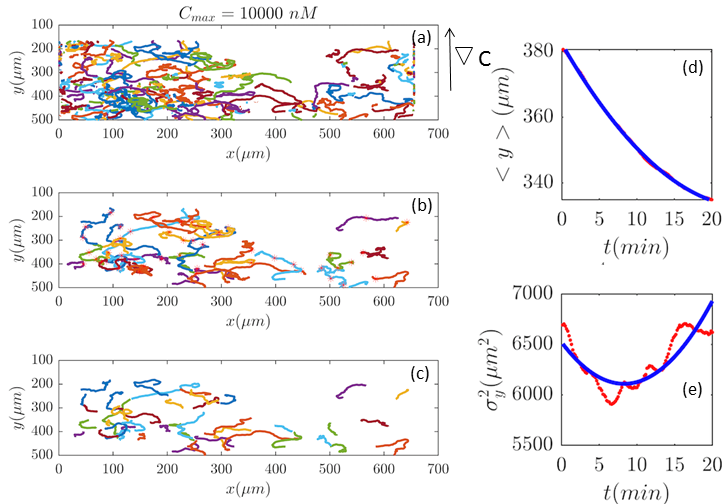}
\caption{(Color online) (a-c) In an experiment with $C_{max}=10000~nM~(\nabla C=28.6~nM/\mu m)$, out of 401 trajectories, 41 are selected and truncated. Comparisons between the experimental measured quantities (red) and the fitted model (blue) are shown in panels (d) and (e).}
\label{fig:track10000nM}
\end{center}
\end{figure*}

\end{document}